\documentclass[preprint,12pt]{elsarticle}

\usepackage{lineno,hyperref,geometry,xcolor,caption}
\usepackage{amsmath,amssymb}
\usepackage{sistyle}
\usepackage[multidot]{grffile}

\journal{Journal Neurocomputing}
\newcommand{\be}{\begin{equation}}
\newcommand{\ee}{\end{equation}}
\newcommand{\beqn}{\begin{eqnarray}}
\newcommand{\eeqn}{\end{eqnarray}}

\begin{document}

\begin{frontmatter}

\title{The effect of noise on the synchronization dynamics of the 
Kuramoto model on a large human connectome graph}

\author[1]{G\'eza \'Odor\corref{cor1}}
\ead{odor@mfa.kfki.hu}
\author[2]{Jeffrey Kelling}
\author[3]{Gustavo Deco}
\address[1]{Institute of Technical Physics and Materials Science,
Centre for Energy Research, P. O. Box 49, H-1525 Budapest, Hungary}
\address[2]{Department of Information Services and Computing,
Helmholtz-Zentrum Dresden-Rossendorf, P.O.Box 51 01 19, 01314 Dresden, Germany}
\address[3]{Center for Brain and Cognition, Theoretical and Computational Group,
Universitat Pompeu Fabra / ICREA, Barcelona, Spain}

\begin{abstract}
We have extended the study of the Kuramoto model with additive Gaussian noise
running on the {\it KKI-18} large human connectome graph. We determined the 
dynamical behavior of this model by solving it numerically in an assumed 
homeostatic state, below the synchronization crossover point we determined 
previously. The de-synchronization duration distributions exhibit power-law
tails, characterized by the exponent in the range $1.1 < \tau_t < 2$, 
overlapping the in vivo human brain activity experiments by Palva et al. 
We show that these scaling results remain valid, by a transformation of the 
ultra-slow eigen-frequencies to Gaussian with unit variance. 
We also compare the connectome results with those, obtained on a regular 
cube with $N=10^6$ nodes, related to the embedding space, and show 
that the quenched internal frequencies themselves can cause frustrated 
synchronization scaling in an extended coupling space.
\end{abstract}


\end{frontmatter}

%%%%%%%%%%%%%%%%%%%%%%%%%%%%%%%%%%%%%%%%%%%%%%%%%%%%%%%%%%%%%%%%%%%%%%%%
\section{Introduction}
%%%%%%%%%%%%%%%%%%%%%%%%%%%%%%%%%%%%%%%%%%%%%%%%%%%%%%%%%%%%%%%%%%%%%%%%

The organization of resting-state activity, i.e. the dynamics of the brain 
under the absence of external stimulation and no task condition, plays 
putatively a critical functional role given the fact that its maintenance 
requires a large part of the total brain’s energy budget~\cite{Atwell01,Raichle06}.  
There are nowadays empirical and computational evidences showing that the 
resting organization facilitates task-based information processing~\cite{FISER2010119}. 
Resting brain networks as captured by Functional Connectivity (FC) maps 
very consistently show task-evoked activity such that individual differences in FC 
can predict individual differences in task-evoked regional 
activity~\cite{Tavor16,Cole16,Osher19}.  
From a mechanistic perspective, whole-brain models were able to demonstrate 
that resting-state organization conforms to a state of ‘criticality’ that 
promotes responsiveness to external stimulation, i.e. resting-state organization 
facilitates task-based processing~\cite{Deco12,66,Senden16}.

Neural activity avalanche measurements found size and duration distributions
that can be fitted by power-laws before a size cutoff, which can arise 
naturally close to a critical point of a second order phase transition
\cite{BP03,Fried,Shew,Yag,brainexp}.
Criticality hypothesis has been advanced, because information processing,
sensitivity, long-range and memory capacity is optimal in the neighborhood
of criticality~\cite{Larr,KC}.
Criticality in models can be defined by the diverging correlation
volume, as we tune a control parameter to a threshold value.

It has been debated how a neural system is tuned to criticality.
At first self-regulatory mechanisms \cite{stas-bak}, leading to 
self-organized criticality \cite{pruessner} were proposed.
Recently, it has been shown that as the consequence of heterogeneity
extended dynamical critical regions emerge in spreading 
models~\cite{MM,HMNcikk} naturally. { 
As real systems are mostly inhomogeneous and one must asses
whether heterogeneity is weak enough to use homogeneous models for
describing them. Heterogeneity can create rare-region 
effects of different relevancy \cite{Vojta2006b}. 
They can generate so-called Griffiths Phases~\cite{Griffiths}
in which scale-free dynamics appears over an extended region around 
the critical point with slowly decaying auto-correlations 
and burstyness \cite{burstcikk}.
This phenomenon was proposed to be the reason for the working
memory in the brain \cite{Johnson}.
Furthermore, in GP the susceptibility is infinite for an entire
range of control parameters near the critical point, providing a
high sensitivity to stimuli, beneficial for information processing.
}

As individual neurons emit periodic signals~\cite{PSM16} it is natural
to expect criticality in oscillator models at the synchronization 
transition point. Very recently analysis of Ginzburg-Landau type 
equations arrived at the conclusion that empirically reported 
scale-invariant avalanches can possibly arise if the cortex is
operated at the edge of a synchronization phase transition, where
neuronal avalanches and incipient oscillations coexist \cite{MunPNAS}.
Several oscillator models have been used in biology, the simplest 
possible one is the Hopf model~\cite{Freyer6353}, which has
been used frequently in neuroscience, as it can describe
a critical point with scale-free avalanches, with sharpened 
frequency response and enhanced input sensitivity.

Indeed, Deco and colleagues developed a mesoscopic whole-brain 
model based on the Hopf model, which provides an excellent account 
of resting-state empirical FMRI~\cite{DKJR} and MEG data~\cite{DECO2017538}. 
Furthermore, the Hopf whole-brain model is able to be used in 
conjunction with higher-resolution functional parcellations that will 
increase model accuracy. The model consists of coupled dynamical units 
representing the cortical and sub-cortical 
brain areas from a given parcellation. The local dynamics of each brain 
area (node) is described by the normal form of a supercritical Hopf 
bifurcation, also called a Landau-Stuart Oscillator, which is the canonical 
model for studying the transition from noisy to oscillatory dynamics. 
The emerging global whole-brain dynamics results from the partial meta-stable 
entrainment of different clusters of brain areas that synchronize and 
build up different network micro-states.

Another complex model, describing more non-linearity%
\footnote{ In the weak coupling limit an equivalence with the 
integrate-and-fire models~\cite{PolRos15} was shown.}
is the Kuramoto model~\cite{kura,Acebron}, with phases $\theta_i(t)$, 
located at $N$ nodes of networks, according to the dynamical equation
\be
\dot{\theta_i}(t) = \omega_{i} + K \sum_{j} W_{ij} 
\sin[ \theta_j(t)- \theta_i(t)] 
\label{kureq}
\ee
The global coupling $K$ is the control parameter of this model, by which
we can tune the system between asynchronous and synchronous states.
The summation is performed over other nodes, with connections described 
by the weighted adjacency matrix $W_{ij}$ and $\omega_{i}$ denotes the 
intrinsic frequency of the $i$-th oscillator. 
For simplicity we used for the $g(\omega_{i})$ distributions 
Gaussian functions~\cite{KurCC}.

Earlier Eq.~(\ref{kureq}) was studied analytically and computationally
on a human connectome graph network of $998$ nodes and in hierarchical
modular networks (HMN), in which moduli exist within moduli in a 
nested way at various scales \cite{Frus}.
As the consequence of quenched, purely topological heterogeneity
an intermediate phase, located between the standard synchronous and
asynchronous phases was found, showing ``frustrated synchronization'',
meta-stability, and chimera-like states~\cite{chimera}.
This complex phase was investigated further in the presence of
noise~\cite{Frus-noise} and on a simplicial complex model of manifolds
with finite and tunable spectral dimension \cite{FrusB} as simple models
for the brain.

We continued to investigate  Eq.~(\ref{kureq}) on a large, weighted human 
connectome network, containing \num{804092} nodes, in an assumed
homeostatic state~\cite{KurCC}. Homeostasis in real brains occurs
via inhibitory neurons~\cite{Homeo-inh,65,66,67,68}, here we modeled
this by normalizing the incoming interaction strengths~\cite{CCdyncikk}.
Recent experiments arrived at a similar conclusion: equalized network 
sensitivity allows critical behavior and produces model results, which
reproduce measured FMRI correlations \cite{Rocha2008}.

Since this graph has a topological dimension $d < 4$~\cite{CCcikk}, 
a real synchronization
phase transition is not possible in the thermodynamic limit, still we could
locate a transition between partially synchronized and desynchronized states.
At this crossover point we observe power-law--tailed synchronization durations,
with $\tau_t \simeq 1.2(1)$, away from experimental values for the brain.
Below the transition of the connectome we found global coupling control-parameter 
dependent exponents $1 < \tau_t \le 2$, overlapping with the range of human 
brain experiments~\cite{brainexp}.%
\footnote{The topological (also called graph) dimension is defined by
\begin{equation} \label{topD}
\langle N_r\rangle \sim r^d \ ,
\end{equation}
where $N_r$ is the number of node pairs that are at a topological
(also called ``chemical'') distance $r$ from each other
(i.e.\ a signal must traverse at least $r$ edges to travel from one
node to the other).} Note however, that in case of the Kuramoto model
the so-called spectral dimension is a more relevant factor influencing
the scaling behavior~\cite{FrusB}.

%%%%%%%%%%%%%%%%%%%%%%%%%%%%%%%%%%%%%%%%%%%%%%%%%%%%%%%%%%%%%%%%%%%%%%%%
\section{Materials and methods}
%%%%%%%%%%%%%%%%%%%%%%%%%%%%%%%%%%%%%%%%%%%%%%%%%%%%%%%%%%%%%%%%%%%%%%%%

Now we extend this study by an additional, annealed zero centered 
Gaussian distributed noise term $\xi(i)$ with unit variance and
a coupling strength $s$
\be
\dot{\theta_i}(t) = \omega_{i} + K \sum_{j} W_{ij}
\sin[ \theta_j(t)- \theta_i(t)] + s \xi(i)
\label{nkureq}
\ee
The effect of noise on synchronization of Kuramoto oscillators
has been investigated by many works (see for example~\cite{10.1143/PTP.79.39}).
It was shown, that in high dimensional, mean-field models the noise can be neglected,
it shifts the critical coupling rate $K_c$ only, but does not change the scaling
behavior. Earlier we found the {\it KKI-18} graph finite dimensional~\cite{CCcikk}
and very heterogeneous, exhibiting a hierarchical modular topology.
In low dimensional heterogeneous systems a large repertoire of attractors can exist,
causing meta-stable states with different degrees of coherence and stability.
Noise enables the system to jump into another close, more stable, attractor.
As neurons work in a noisy background this study can be interesting for neuroscience.

We follow the dynamical behavior of the system through studying the Kuramoto 
order parameter defined by
\be
R(t)=\frac{1}{N}\left|\sum_{j=1}^Ne^{i\theta_j(t)}\right|,
\label{op}
\ee
which is finite, above a critical coupling strength $K > K_c$, or tends to 
$O(N^{-1/2})$ for $K < K_c$. At $K_c$, in case of an incoherent initial state
it evolves as
\be
R(t,N) = N^{-1/2} t^{\eta} f_{\uparrow}(t / N^{\tilde z}) \ ,
\label{escal}
\ee
characterized by the dynamical exponents: ${\tilde z}$, $\eta$ 
and $f_{\uparrow}$ denotes a scaling function.

Within the framework of the synchronization model we identified a spontaneous 
avalanche start times at $t=0$ of the fully desynchronized initial condition 
and the end times $t_x$, when $R(t)$ returned back to $1/\sqrt(N)$, 
related to the synchronization value of independent oscillators.
Note, that by changing the start times to the first up crossing value
did not change the tail behavior we consider here. To obtain the avalanche
duration probability distributions $P(t_x)$ we performed $\simeq 10^4$ runs, 
with independent random $\omega_{i}$ intrinsic frequencies and applied 
histogramming method with increasing bin sizes: $\Delta t_x \propto t_x^{1.12}$.

The following graphs have been considered:
\begin{enumerate}
\item 3D lattice with linear size $L=100$ and periodic boundary conditions.
\item Weighted, symmetric large human connectome graph: {\it KKI-18} 
\cite{CCcikk} downloaded from the Open-connectome project \cite{OCP}. 
\end{enumerate}

To integrate the differential equation (\ref{kureq}) we used a Graphics
card (GPU) code, 
based on the fourth order Runge-Kutta method of Numerical Recipes~\cite{NumR} 
and the boost  library odeint~\cite{boostOdeInt} on various networks. 
Step sizes: $\Delta = 0.1, 0.01, 0.001$ have been tested and finally the 
$\Delta =0.01$ precision found to be sufficient, for $s \le 2$.
For larger noise amplitudes $s$ this precision was not enough.
In case of our initial attempt, with natural $\omega_{i} < 0.02$ we needed
large $s$ values to see a synchronization transition and even $\Delta < 0.01$
turned out to be insufficient, making the numerical analysis
prohibitively time-consuming for available computer resources.

In general, the $\Delta < 0.01$ precision did not improve the stability of the
solutions further, but caused smaller fluctuations due to the chaotic 
behavior of Eq.~(\ref{kureq}), which could be compensated by averages over 
many independent samples with different $\omega_{i}$.
We used the $\epsilon = 10^{-12}$ criterion in the Runge-Kutta algorithm and
parallelized the solver for NVIDIA graphic cards, by which we could
achieve a $\sim\times 40$ increase in the throughput on Tesla V100 GPU-s
with respect to a single $12$-core Intel Xeon Gold 6136 CPU.
Algorithmic and benchmark details will be discussed elsewhere \cite{GPU-kur}.

We determined the Kuramoto order parameter at a fixed $K$, by increasing the 
sampling time steps exponentially
\begin{equation}
t_k = 1 + 1.08^{k} \ ,
\end{equation}
which is a common method at critical systems, where we expect power-law (PL) 
asymptotic time dependences. We estimated $t_x = (t_k + t_{k-1})/2$, 
where $t_k$ was the first measured down crossing time.
The initial conditions were generally $\theta_i(0) \in (0,2\pi ]$ phases, 
with uniform distribution, describing fully disordered states. 
The probability distribution tails were fitted using the least squares 
fit method beyond a time, fixed by visual inspection of the results.
To visualize corrections to PL scaling we determined the effective exponents
of $R$ as the discretized, logarithmic derivative of Eq.~(\ref{escal})
\begin{equation}  \label{Reff}
\eta_\mathrm{eff} = \frac {\ln \langle R(t_{k+3})\rangle - \ln \langle R(t_{k})\rangle} 
{\ln(t_{k+3}) - \ln(t_{k})} \ ,
\end{equation}
were the brackets denote sample averaging over different initial conditions.

\begin{figure}[!h]
\includegraphics[height=5.5cm]{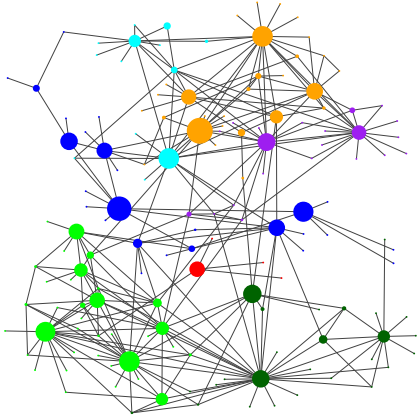}
\caption{Network of the modules of the {\it KKI-18} human connectome graph.
 The size of circles is proportional with the number of nodes. The network of
 modules itself shows modularity arising from the hierarchical structure of the
 {\it KKI-18} connectome. The each circle's color indicates its membership in
 one of seven modules obtained through Leiden community analysis of the
 displayed cluster graph.
 }
\label{modules}
\end{figure}

We obtained the {\it KKI-18} graph from the Open Connectome project repository~\cite{OCP}. 
This is based on the diffusion tensor image~\cite{DTI}, approximating the 
{\it structural connectivity} of the white matter of a human brain.
The graph version we downloaded in 2015 comprises a large component with
$N = \num{804092}$ nodes, connected via \num{41523908} undirected edges 
and several small sub-components, which were ignored here
~\footnote{Note, that keeping the sub-components, did not change the results 
within numerical accuracy.}.
This graph allowed us to run extensive dynamical studies on present day
CPU/GPU clusters, large enough to draw conclusions on 
the scaling behavior without too strong finite size effects, hindering
scaling regions.
The large connectomes from~\cite{OCP} of the human brain possess 
\SI1{mm^3} resolution, obtained by a combination of diffusion weighted, 
functional and structural magnetic resonance imaging scans. 
These are symmetric, weighted networks, where the weights measure the 
number of fiber tracts between nodes.
The {\it KKI-18} graph is generated via the MIGRAINE method,
described in~\cite{MIG}. 
It exhibits a hierarchical modular structure by the construction from the 
Desikan cerebral regions with (at least) two quite different scales. 
The graph topology is displayed on Fig.~\ref{modules}, in which modules 
were identified by the Leiden algorithm~\cite{TraagWaltmaVanEck2018_LouvainLeiden},
and the network of modules generated and visualized using the python-igraph
library~\cite{igraph}. This identified $153$ modules, with sizes varying between 
$7$ and \num{35332} nodes in the displayed case, however since this is a
heuristic approach, these numbers vary by about \SI{10}{\%}.
A recent experimental study has provided confirmation for the connectome generation used here
~\cite{newcomp}. This suggests that diffusion MRI tractography is a powerful tool for 
exploring the structural connection architecture of the brain.

In~\cite{CCcikk} it was found that, contrary to the small world network
coefficients, these graphs exhibit topological dimension $D=3.05$~\cite{CCcikk}, 
slightly above the embedding space and a certain amount of universality, 
supporting the selection of
\textit{KKI-18} as a representative of the large human connectomes available.

To keep a local sustained activity requirement for the brain \cite{KH} 
and to provide a homeostatic state, we modified  {\it KKI-18} by normalizing the
incoming weights of node $i$ in \cite{CCdyncikk}:
$W'_{i,j} = W_{i,j}/\sum_{j \in {\rm neighb. of} \ i} W_{i,j}$
at the beginning of the simulations. In reality such local homeostasis is
the consequence of the competition of exhibitory and inhibitory neurons.

%%%%%%%%%%%%%%%%%%%%%%%%%%%%%%%%%%%%%%%%%%%%%%%%%%%%%%%%%%%%%%%%%%%%%%%%
\section{Results}
%%%%%%%%%%%%%%%%%%%%%%%%%%%%%%%%%%%%%%%%%%%%%%%%%%%%%%%%%%%%%%%%%%%%%%%%

\subsection{The 3D lattice}

In \cite{KurCC} we compared the connectome results with small-world
graphs, generated from 2D lattices with additional random long-range
links (2Dll). This was done as both the connectome and the 2Dll graphs
exhibit small world properties and quenched topological disorder on
top of the intrinsic heterogeneity.
As the topological dimension of the connectome is slightly above the
$D=3$ embedding space~\cite{CCcikk}, due to the long connections, 
now we have studied the growth of $R(t)$ on the 3D lattice of linear 
size $L=100$ by starting from states of oscillators with fully random phases
and by averaging over $\num{5000}-\num{10000}$ internal frequency realizations 
up to $t = 10^3$ time steps. In this case we can test the effect of 
quenched heterogeneity of the $\omega_i$-s on the dynamical
behavior, using a zero centered, unit variance $g(\omega)$ distribution.
 
\begin{figure}[!h]
\includegraphics[height=5.5cm]{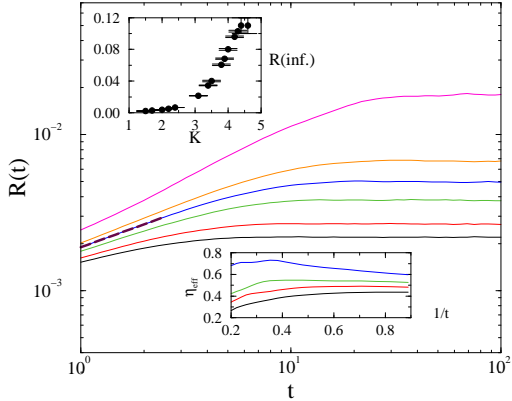}
\caption{Growth of the average $R$ on the 3D lattice for Gaussian
quenched $\omega_i$ and couplings $K=1.5, 1.7, 2.0, 2.2, 2.4, 3.0$ 
(bottom to top curves). The lower inset shows the effective exponents
defined as (\ref{Reff}). The upper inset shows the steady state values
as the function of $K$.}
\label{fig3D}
\end{figure}

Fig~\ref{fig3D} shows how the synchronization order parameter grows 
below the transition. Note, that for comparison with the connectome
simulations we used incoming weight normalization, which means a
constant factor $6$, the incoming degree in 3D, by which the global
couplings are renormalized with respect to the original (\ref{kureq}).
The initial growth changes from convex to concave at $K_c \simeq 2.2$,
suggesting a crossover point there. This value is higher than what we obtained for the
{\it KKI-18} connectome~\cite{KurCC}: $K_c=1.7$, as we have much lower connectivity
now. The {\it KKI-18} graph has an average node degree: $\langle k\rangle=156$,
in contrast with the cubic lattice: $\langle k\rangle=6$,
thus we need stronger global coupling to achieve synchronization.
The initial behavior of growth at $K_c=2.2$ is like $R(t) \propto t^{0.5}$,
with an exponent somewhat smaller, but close to the estimates obtained for 
the 2Dll and for the {\it KKI-18} graphs: $\eta=0.6(1)$. This can be read-off
from the local slopes before finite-size cutoff, shown in the inset of
Fig~\ref{fig3D}. The upper inset of Fig~\ref{fig3D} shows the steady 
state values of $R(t)$, obtained by averaging over the realizations 
for $t > 50$.

\begin{figure}[!h]
\includegraphics[height=5.5cm]{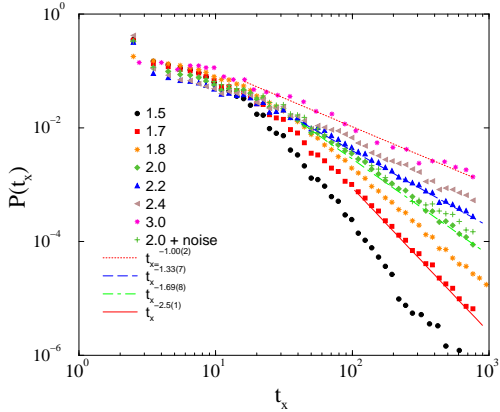}
\caption{Duration distribution of $t_x$ on 3D lattice for 
couplings: $K=1.5$ (bullets), $1.7$ (boxes),  $1.8$ (stars), $2.0$ (diamonds), 
$2.2$ (up triangles), $2.2$ (right triangles), $3.0$ (stars). 
Lines show a PL fits for the tails. Noisy case: $K=2.0$ (plus sign).}
\label{dur3d}
\end{figure}

As for the connectome we determined the first crossing times $t_x$,
when $R(t)$ of single realizations first fell below the threshold
value $1/\sqrt(N) = 0.001$.
Following the usual histogramming procedure we obtained the de-synchronization 
distributions, which can be considered as avalanches, induced by a 
spontaneous synchronization and a relaxation process.
The tail of the relaxation time distribution PDF at $K_c=2.2$ decays as 
$\simeq t_x^{-1.33(7)}$ (see Fig.~\ref{dur3d}), obtained by least 
squares power-law regression, applied for the tail region $t_x > 30$.
Below $K_c$ the curves shown on the figure provide good agreement 
with non-universal PL-s. However, for $K < 1.7$ the PL region seems to shrink.
The ineffectiveness of the weak additive noise has also been demonstrated in case
$K=2.0$ coupling, using a Gaussian annealed term of strength $s=1$.

{
The emergence of frustrated synchronization by $\omega_i$ heterogeneity
is similar to the contact process with site disorder~\cite{Vojta2006b},
where Griffiths Phase has been found in various spatial dimensions.
To provide a counter-example we have also studied the 3D case with 
uniform oscillators and annealed noise. In the Appendix we show that
in this case no extended scaling region, but a single critical point
emerges.
}

%%%%%%%%%%%%%%%%%%%%%%%%%%%%%%%%%%%%%%%%%%%%%%%%%%%%%%%%%%%%%%%%%%%%%%%%
\subsection{The Connectome graph}
%%%%%%%%%%%%%%%%%%%%%%%%%%%%%%%%%%%%%%%%%%%%%%%%%%%%%%%%%%%%%%%%%%%%%%%%

\subsubsection*{Mapping ultra-slow self-frequency scales onto computationally
feasible oscillations}

FMRI measurements~\cite{Ponce-Deco-Plos15,DKJR} found that in the human brain 
global phase synchrony of the BOLD signals evolves on a characteristic 
ultra-slow: $<0.01$ Hz time scale. 
In modeling this, on smaller sized connectomes, using the noisy Kuramoto 
equation~\cite{Ponce-Deco-Plos15} and the Hopf model~\cite{DKJR}, the intrinsic 
frequencies were filtered in the $0.04 < \omega < 0.07$ Hz band.
In both cases the temporal and spatial synchronization patterns were found to 
approximate well the empirical data.
Here we extend this modeling for a large human connectome, which allows to
test possible scaling behavior, with $\omega_i$-s of narrow spread ($\sigma=0.02$) 
and typical mean value $\overline\omega = \langle \omega_i \rangle = 0.05$.

First we show, that in the case of the Kuramoto model the narrow frequency band 
can be mapped onto the much wider $\sigma=1$, $\overline\omega=0$ 
Gaussian case, where we obtained recent results~\cite{KurCC} on synchronization
durations.
The invariance over the mean frequency is obvious, as (\ref{kureq}) 
is invariant to the global shift of a mean rotation frame 
$\overline\omega \to \overline\omega'$.
The oscillation-size dependence can also be gauged out by the following
transformation: $\omega_i \to a \omega_i'$, $t \to (1/a)t'$ and $K\to a K'$.
Therefore, for small $a$ values corresponding to empirical data,
we can obtain the same results as for $\sigma=1$ at rescaled, late times 
and using small global couplings.

From a technical point of view this is very important, because we can
recover the asymptotic results at ultra-slow frequencies using much 
shorter time scales in our simulations.
Note, that in the case of the Hopf model the rotating frame transformation
is also possible, but the width of the $g(\omega)$ distribution cannot be gauged
out and acts as an independent parameter, causing a nontrivial phase 
structure of the oscillations~\cite{MS90}.
Furthermore, a completely band filtered $g(\omega_i)$ spectrum with a 
flat top can be transformed onto the uniformly distributed natural frequencies. 
In this case it is known, that in the limit of infinite system size
the Kuramoto model in the steady state undergoes a first-order phase 
transition~\cite{PhysRevE.72.046211}.
At first-order transitions we don't expect critical scaling behavior and indeed 
our simulations for this case do not show PL de-synchronization duration tails
(see Appendix).

%%%%%%%%%%%%%%%%%%%%%%%%%%%%%%%%%%%%%%%%%%%%%%%%%%%%%%%%%%%%%%%%%%%%%%%%
\subsubsection{Noisy Kuramoto results}
%%%%%%%%%%%%%%%%%%%%%%%%%%%%%%%%%%%%%%%%%%%%%%%%%%%%%%%%%%%%%%%%%%%%%%%%

Having taken into account the transformation properties of the previous section 
we investigated the effect of annealed noise, by adding a time varying Gaussian 
distributed random numbers to the (\ref{kureq}). 
Earlier in~\cite{KurCC} we concluded that using step sizes $\Delta < 0.1$
in the Runge-Kutta-4 solver did not modify the results. 
Here we test the effect of $\Delta$ again, as we add a stochastic noise 
on top of the chaoticity, inherent in the noiseless Kuramoto equation.

\begin{figure}[!h]
 \includegraphics[height=5.5cm]{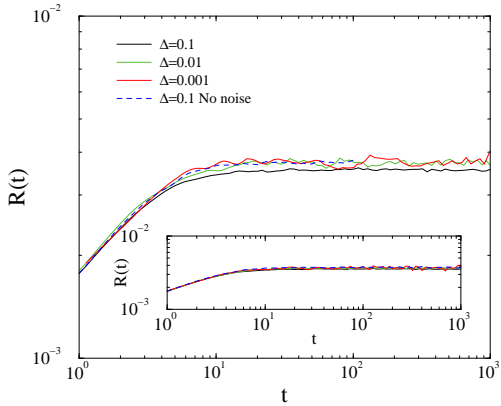}
\caption{Growth of the average $R$ on the {\it KKI-18} graph below the synchronization
transition point at $K=1.4$ for $s=1$, using different precision: $\Delta=0.1, 0.01, 0.001$. The dashed line shows the noiseless case result obtained by  $\Delta=0.1$.
The inset shows the same for $s=2$.}
\label{figccgrowth}
\end{figure}

First we applied $s=1$ noise to the {\it KKI-18} homeostatic graph case~\cite{KurCC}.  
As Fig.~\ref{figccgrowth} shows slightly below the synchronization
transition point: $K=1.4 < K_c=1.7$ the Kuramoto order parameter growth 
curves fully agree with the noiseless case~\cite{KurCC} one and the applied
$\Delta=0.1, 0.01, 0.001$ step sizes do not affect too much the growth
regime. One can see a smaller saturation value for $\Delta=0.1$ than for
$\Delta = 0.01, 0.001$.
Here each line corresponds to an average over $\sim 10^4$ realizations,
in which both the quenched and the annealed noise varies. 

\begin{figure}[!h]
 \includegraphics[height=5.5cm]{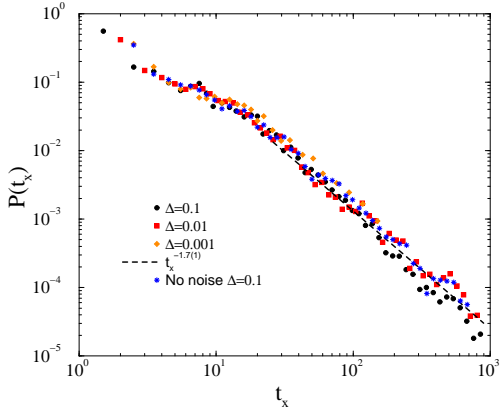}
\caption{Duration distribution of $t_x$ on the {\it KKI-18} model 
at $K=1.4$ for $s=1$ using different precisions: $\Delta=0.1$ (bullets), 
$\Delta=0.01$ (boxes),  $\Delta=0.001$ (diamonds).
The dashed line shows a PL fit for the tail region: $t_x > 10$ of
the $\Delta=0.01$ data. The stars show former results, obtained for the
noiseless case.}
\label{durcc}
\end{figure}

We have determined the first crossing times $t_x$, when $R$ fell below: 
$1/\sqrt(N) = 0.001094$.
Following the histogramming procedure we obtained the distributions 
$p(t_x)$, which again show PL tails, characterized by exponents:  
$\tau_t=1.2(1)$ (see Fig.~\ref{durcc}). 
The fitted $\tau_t=1.9(1)$ exponent is in the range of in vivo
human neuro experiments: $1.5 < \tau_t < 2.4$ \cite{brainexp}. 
Then we repeated the analysis for $K=1.3$,  $s=1$ and found similar
agreement with the noiseless results (see figure in the Appendix).

\begin{figure}[!h]
\includegraphics[height=5.5cm]{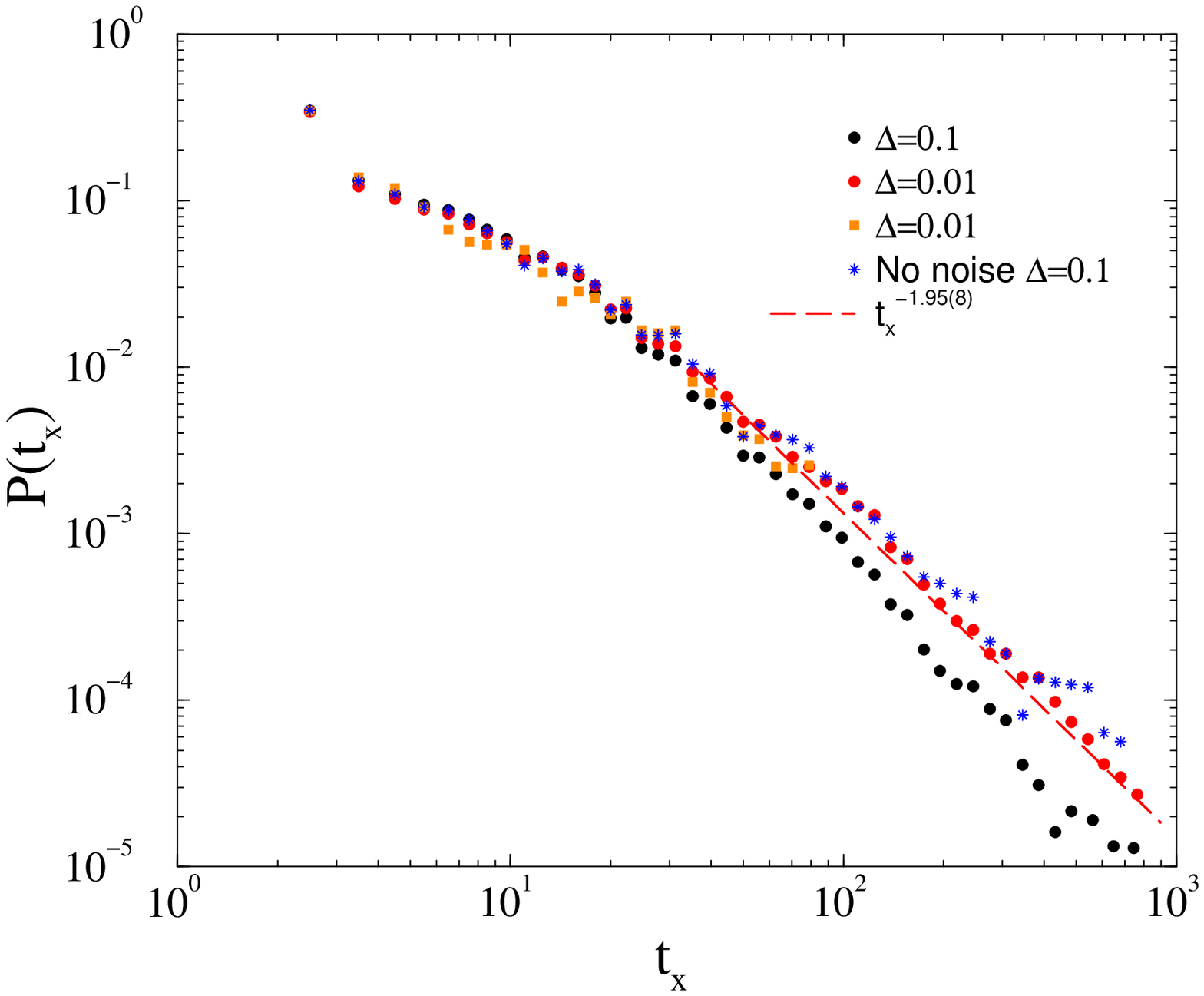}
\caption{Duration distribution of $t_x$ on the {\it KKI-18} model
at $K=1.4$ for $s=2$ using different precisions: $\Delta=0.1$ (bullets),
$\Delta=0.01$ (boxes),  $\Delta=0.001$ (diamonds).
The dashed line shows a PL fit for the tail region: $t_x > 10$ of
the $\Delta=0.01$ data. The stars show former results, obtained for the
noiseless case.}
\label{durcc2}
\end{figure}

We have tested noise: $s=2$ at $K=1.4$ as shown on the inset of 
Fig.~\ref{figccgrowth} and on \ref{durcc2}, but again we just found insensitivity
in the growth and scaling results. We have learned, that going to even 
stronger noise amplitudes requires smaller $\Delta$-s to achieve numerical
precision independence and it is questionable from neuroscience point
of view if it is worth to study such strong noises. We plan to study
this later.

\begin{figure}[!h]
\includegraphics[height=5.5cm]{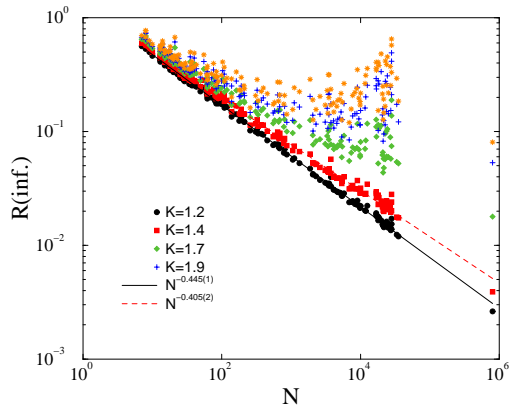}
\caption{Size dependence of the steady state values of~$R$ in the modules at
different couplings~$K$. One can observe normal finite-size scaling below
the transition and breaking of this at or above the transition point.
The single points to the far right for each series correspond to steady state
values $R(\mathrm{inf.})$ for the whole graph. The lines show PL fits below
the transition point.
}
\label{RcomSatMulti}
\end{figure}

{
In order to see the effect of the modules on the synchronization behavior 
in greater detail, we also computed the order parameter for each module 
separately. Fig.~\ref{RcomSatMulti} shows the steady-state values for 
each module for different coupling parameters. While normal finite-size 
scaling can be observed below the transition point, larger modules start 
to synchronize above. We fitted PL-s for the sub-critical scaling, showing
non-universal exponents in an extended $K$ region.}

%%%%%%%%%%%%%%%%%%%%%%%%%%%%%%%%%%%%%%%%%%%%%%%%%%%%%%%%%%%%%%%%%%%%%%%%
\section{Conclusion and discussion}
%%%%%%%%%%%%%%%%%%%%%%%%%%%%%%%%%%%%%%%%%%%%%%%%%%%%%%%%%%%%%%%%%%%%%%%%

Resting-state activity of the brain can be modeled by simple
whole-brain models, exhibiting critical behavior at the edge of the
synchronization transition. 
We have investigated the dynamical synchronization behavior of the
Kuramoto model on a large, weighted human connectome network.
{The dynamical behavior of heterogeneous Kuramoto, especially for local
interactions is a largely unexplored field, according to out knowledge.
In case of identical oscillators heterogeneous phase lags or couplings
have been shown to result in partial synchronization and stable chimera
states~\cite{Feng_2015,PhysRevLett.101.264103,LAING20091569,PhysRevE.89.022914}
Realistic models of the brain, however, require oscillators~\cite{CABRAL2011130}
to be heterogeneous.}

In particular, we focused on the effect of additive noise on the
de-synchronization duration distributions. In Ref.~\cite{KurCC}
it was found that below the phase synchronization transition point
these distributions exhibit non-universal power-law tails, with exponents
overlapping with the empirical activity avalanche duration values
for humans in vivo.  

We found that weak Gaussian noises with amplitudes not larger than
those of the quenched Gaussian self-frequencies do not affect the
previous results within numerical precision. This means that
time-dependent, thermal like noise does not destroy or alter the
dynamical scaling behavior of this model.

Ref.~\cite{Frus} studied the synchronization behavior of the Kuramoto
model on hierarchical modular networks with $N=4096$ and human connectomes
with $N=998$ nodes.
They found stretched exponential decay tails for the order parameter
$\rho(t) = 1 - \langle R(t) \rangle$ in the absence of frequency
heterogeneity, in agreement with analytic approximations using
$\omega_{i}=0$. { They assumed that fixing all intrinsic frequencies 
to be identical does not decrease generality of the results.}
We have also tried to fit our $p(t_x)$ results with stretched exponential 
functions, (see Appendix) but reasonable agreement could be found
only for very low couplings, far below $K_c$. So, we conclude that
that the quenched disorder in the self-frequencies cause PL tails
in the dynamical behavior of chimera-like states at the
edge of criticality. These non-universal PL-s resemble to Griffiths
Phase effects and the results obtained in case of the second order 
Kuramoto model for power-grids below the synchronization
transition~\cite{POWcikk}. { We have also detected module size 
scaling of $R(t\to\infty)$ below the transition point, with 
$K$-dependent exponents, which would be an interesting subject
of further study.}

We have shown that the empirical results with ultra-slow oscillations 
can be transformed onto zero mean Gaussian frequencies as the consequence
of the Galilean symmetry of the Kuramoto equation.
As we used the 4th order Runge-Kutta solver, we carefully confirmed that
the step size $\Delta=0.01$ is sufficient for the increased precision
required by the additive noise. This sensitivity is related to the
fast changes of the time dependent, additive noise.

Another point is the positiveness of the $g(\omega_{i})$ distribution in
the brain, as we don't expect neural oscillators 'rotating backwards'.
This corresponds to the question of asymmetric distribution of natural
frequencies, such that for $g(\omega_{i}) = 0$ for $\omega_{i} < 0$.
It has been shown that in case of uni-modal $g(\omega_{i})$-s
only the first derivative, the flatness of $g(\omega_{i})$, matters
here. Without a flat top, like an asymmetric triangle, one obtains the
same universal critical behavior ($\beta=1/2$) as for the original Kuramoto
model with zero centered Gaussian~\cite{BU08}. Thus we expect the same dynamical
behavior for an asymmetric, truncated Gaussian:
$g(\omega_{i}) = 0$ for $\omega_{i} < 0$ as we found by the symmetric Gaussian.

We have compared the results with those obtained on regular 3D lattices
of similar size and found, that the heterogeneity of the self-frequencies
already generate the non-universal scaling region, which can be called
a frustrated synchronization phase, exhibiting chimeras~\cite{chimera}.
{At the synchronization transition point we found: $\tau_t = 1.33(7)$, 
slightly higher than in case of the connectome:
$\tau_t = 1.2(1)$~\cite{KurCC} and well below the mean-field value:
$\tau_t = 1.6(1)$~\cite{KurCC}, so the topology plays a role mainly
via the graph dimension. Determination of these exponent estimates 
depend strongly on the precise location of $K_c$, which is harder 
to get in smaller systems, suffering finite-size cutoffs. 
Therefore the large systems considered here are justified. 
The local, 3D connected results can also be important for neuroscience, as
local cortical connections appear before more distant ones at the creation
of the circuit~\cite{Tau-pet-10}. Thus this study can be interesting 
for understanding young brains, containing local connections mainly.
Furthermore, 3D networks may occur in case of artificial intelligence 
neural systems.}

The dynamical scaling behaviors have been found to be robust, supporting
universality even if the Kuramoto model could be considered too 
simplistic to describe the brain. Note, however that in the weak-coupling
limit equivalence of phase-oscillator and integrate-and-fire models was
found \cite{PolRos15}, which may hold for the sub-critical region,
where we observed the dynamical scaling.
This provides a support for the edge-of-criticality hypothesis of
oscillating systems near and below the synchronization transition
point.

An interesting continuation of this work would be the study of
the effect of phase shifts, caused by the finite signal propagation
in the neural network { or the introduction of a threshold, as in 
integrate-and-fire models, although by universality of critical 
systems we don't expect qualitative change in the scaling behavior.}

%%%%%%%%%%%%%%%%%%%%%%%%%%%%%%%%%%%%%%%%%%%%%%%%%%%%%%%%%%%%%%%%%%%%%%%%
\section*{Appendix}
%%%%%%%%%%%%%%%%%%%%%%%%%%%%%%%%%%%%%%%%%%%%%%%%%%%%%%%%%%%%%%%%%%%%%%%%

\setcounter{figure}{0} \renewcommand{\thefigure}{A.\arabic{figure}}

\begin{figure}[!h]
\includegraphics[height=5.5cm]{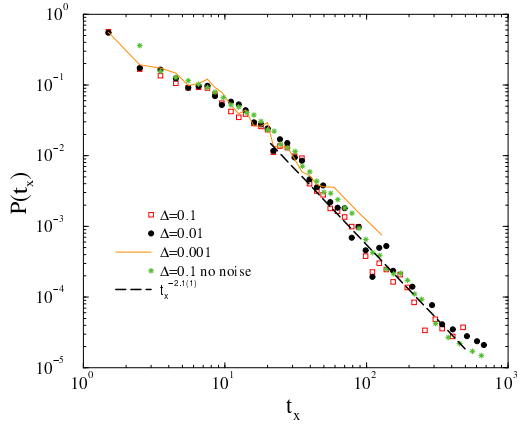}
\caption{Duration distribution of $t_x$ on the {\it KKI-18} model
at $K=1.3$ for $s = 1$ using different precisions: $\Delta=0.1$ (boxes),
$\Delta=0.01$ (bullets),  $\Delta=0.001$ (thin line).
The dashed line shows a PL fit for the tail region: $t_x > 20$ of
the $\Delta=0.01$ data. The stars show former results, obtained for the
noiseless case.}
\label{A1}
\end{figure}

In this Appendix we show results for the synchronization
duration distribution of Kuramoto on the {\it KKI-18} model at $K=1.3$.
As Fig.~\ref{A1} shows dependence on the numerical integration step size
$\Delta$ is negligible and the results agree with the former noiseless
case.
A least-squares fit for the tails results in $\tau_t=2.1(1)$, which is in
the experimentally measured range for activity avalanche durations.

\begin{figure}[!h]
\includegraphics[height=5.5cm]{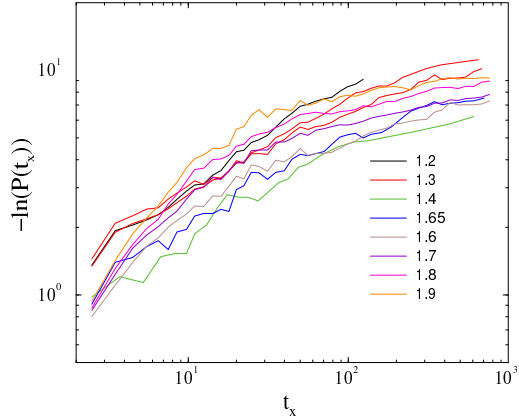}
\caption{Duration distribution of $t_x$ on the {\it KKI-18} model
for different $K$ values shown in the legends.}
\label{A2}
\end{figure}

We have also tried to test in the noiseless case whether the
$P(t_x)$ distributions would follow stretched-exponential decay
instead of PL-s. As Fig.~\ref{A2} shows the $-\ln(P(t_x)))$ curves do not
exhibit straight lines on the log-log plot, except perhaps for $K=1.2$ only,
which is far from $K_c$.

\begin{figure}[!h]
\includegraphics[height=5.5cm]{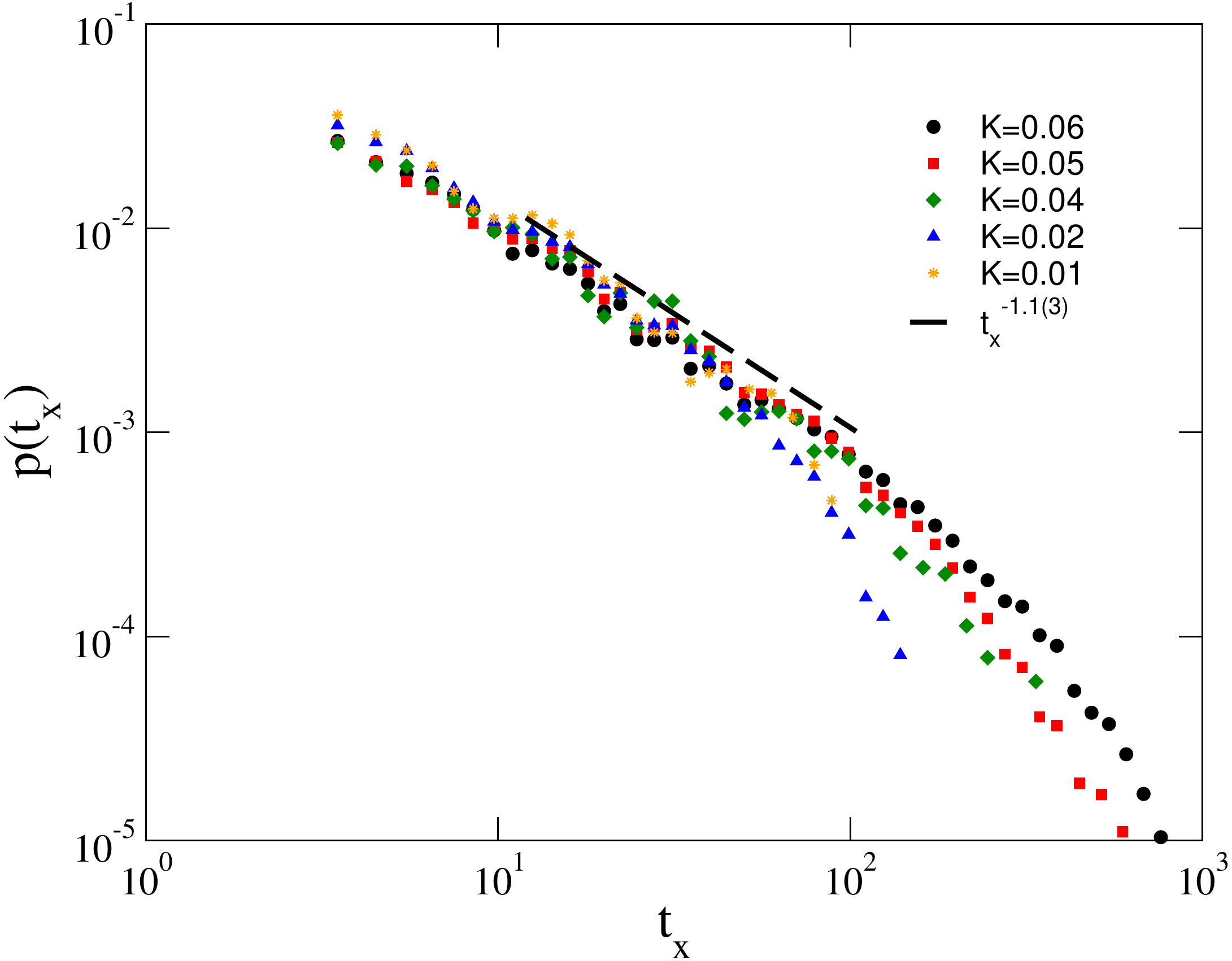}
\caption{Duration distribution of $t_x$ in the homogeneous model
Kuramto model for different $K$ values shown in the legends.
At the critical point: $K_c=0.05(1)$ we see XY model-A dynamical behavior,
and exponential decays sub-critically.}
\label{A3}
\end{figure}

The method has been tested by considering 3D lattices of size $L=100$.
of uniform oscillators and annealed noise with $s = 5$ amplitude.
In this case we don't have any heterogeneity and an order-disorder phase
transition emerges by increasing the coupling, which should belong to the
XY criticality in 3D~\cite{odorbook}. 
As one can see on the attached graph (Fig.~\ref{A3}) we obtain a PL at 
the phase transition point $K = 0.05(1)$ and fast decays below it. 
Here the fitted slope is $\tau_t = 1.1(3)$. 
This agrees with the expectation, coming from scaling 
relations and known data for the critical dynamics of the XY with
model-A dynamics in 3D: $\beta=0.347(1)$, $\nu_{\perp}=0.670(1)$~\cite{PELISSETTO2002549},
$Z=2$~\cite{3DXYZ}, $\tau_t = 1 + \beta / (\nu_{\perp} Z) = 1.25(2)$.
Note, that for $K > 0.05$ the distributions become singular, 
characterized by $1/t_x$ decay, before a finite size cutoff.

\begin{figure}[!h]
\includegraphics[height=5.5cm]{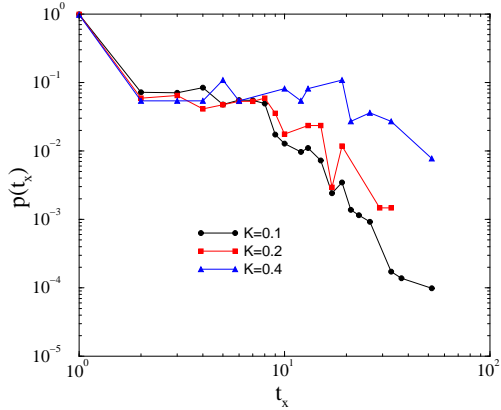}
\caption{Duration distribution of $t_x$ in case of Kuramto with uniformly
distributed $g(\omega_i)$ for different $K$ values shown in the legends.
One cannot observe PL-s here.}
\label{A4}
\end{figure}

Here we also show duration distribution results for the Kuramoto model 
with uniform $g(\omega_i)$, which undergoes a first-order phase transition.
We do not expect critical scaling behavior and indeed our simulations for 
this case do not show PL de-synchronization duration tails.
The $p(t_x)$ curves saturate, followed by a sharp decay for different K-s
as shown on Fig.~\ref{A4}.

\section{Abbreviations}
GPU, Graphics Processing Unit; CPU Central Processing Unit; 
FMRI, Functional magnetic resonance imaging; MEG Magnetoencephalography;
PDF, Probability distribution function; PL, power-law; KKI, Kennedy Krieger Institute;
HMN, hierarchical modular network; MRI, diffusion magnetic resonance imaging;
2Dll, 2D lattices with additional random long-range links;
BOLD, Blood-oxygen-level-dependent imaging.

\section{Competing interest}
The authors declare that they have no competing interests.

\section{Consent for publication}
Not applicable.

\section{Ethics approval and consent to participate}
Not applicable.

\section{Funding}
The work has been performed under the Project HPC-EUROPA3 (INFRAIA-2016-1-730897),
with the support of the EC Research Innovation Action under the H2020 Programme;
in particular, G. \'O. gratefully acknowledges the support of Center for Brain and
Cognition Theoretical and Computational Group Universitat Pompeu Fabra / ICREA Barcelona
and the computer resources and technical support provided by BSC Barcelona.
Support from the Hungarian National Research, Development and Innovation
Office NKFIH (K128989), the Initiative and Networking Fund of the Helmholtz
Association via the W2/W3 Programme \mbox{(W2/W3-026)} and the Helmholtz
Excellence Network DCM-MatDNA (ExNet-0028) is acknowledged.

\section{Availability of data and materials}
The codes and the graphs used here are available on request from the
corresponding author.

\section{Authors' contributions}
G. \'O designed, coordinated this research, drafted the manuscript
and performed computations and data analysis.
J. K. developed GPU codes and participated in data analysis.
G. D. contributed to the manuscript, conceivied of the study and 
participated in research coodination.
The authors read and approved the final manuscript.

%%%%%%%%%%%%%%%%%%%%%%%%%%%%%%%%%%%%%%%%%%%%%%%%%%%%%%%%%%%%%%%%%%%%%%%%
\section*{Acknowledgments}
%%%%%%%%%%%%%%%%%%%%%%%%%%%%%%%%%%%%%%%%%%%%%%%%%%%%%%%%%%%%%%%%%%%%%%%%

We thank R. Juh\'asz and A. Ponce for the useful comments and Wesley Cota for
generating and providing us modules of KKI-18 as well as precursor of
Fig.~\ref{modules}.
We gratefully acknowledge computational resources provided by, NIIF Hungary, 
the HZDR computing center, the Group of M. Bussmann and the Center 
for Information Services and High Performance Computing (ZIH) at TU Dresden 
via the GPU Center of Excellence Dresden. We thank S. Gemming for support.

%%%%%%%%%%%%%%%%%%%%%%%%%%%%%%%%%%%%%%%%%%%%%%%%%%%%%%%%%%%%%%%%%%%%%%%%
\bibliography{bib}

\begin{thebibliography}{10}
\expandafter\ifx\csname url\endcsname\relax
  \def\url#1{\texttt{#1}}\fi
\expandafter\ifx\csname urlprefix\endcsname\relax\def\urlprefix{URL }\fi
\expandafter\ifx\csname href\endcsname\relax
  \def\href#1#2{#2} \def\path#1{#1}\fi

\bibitem{Atwell01}
D.~Attwell, S.~B. Laughlin,
  \href{https://doi.org/10.1097/00004647-200110000-00001}{An energy budget for
  signaling in the grey matter of the brain}, Journal of Cerebral Blood Flow \&
  Metabolism 21~(10) (2001) 1133--1145, pMID: 11598490.
\newblock \href
  {http://arxiv.org/abs/https://doi.org/10.1097/00004647-200110000-00001}
  {\path{arXiv:https://doi.org/10.1097/00004647-200110000-00001}}, \href
  {https://doi.org/10.1097/00004647-200110000-00001}
  {\path{doi:10.1097/00004647-200110000-00001}}.
\newline\urlprefix\url{https://doi.org/10.1097/00004647-200110000-00001}

\bibitem{Raichle06}
M.~E. Raichle, \href{https://science.sciencemag.org/content/314/5803/1249}{The
  brain{\textquoteright}s dark energy}, Science 314~(5803) (2006) 1249--1250.
\newblock \href
  {http://arxiv.org/abs/https://science.sciencemag.org/content/314/5803/1249.full.pdf}
  {\path{arXiv:https://science.sciencemag.org/content/314/5803/1249.full.pdf}},
  \href {https://doi.org/10.1126/science. 1134405} {\path{doi:10.1126/science.
  1134405}}.
\newline\urlprefix\url{https://science.sciencemag.org/content/314/5803/1249}

\bibitem{FISER2010119}
J.~Fiser, P.~Berkes, G.~Orban, M.~Lengyel,
  \href{http://www.sciencedirect.com/science/article/pii/S1364661310000045}{Statistically
  optimal perception and learning: from behavior to neural representations},
  Trends in Cognitive Sciences 14~(3) (2010) 119--130.
\newblock \href {https://doi.org/https://doi.org/10.1016/j.tics.2010.01.003}
  {\path{doi:https://doi.org/10.1016/j.tics.2010.01.003}}.
\newline\urlprefix\url{http://www.sciencedirect.com/science/article/pii/S1364661310000045}

\bibitem{Tavor16}
I.~Tavor, O.~P. Jones, R.~B. Mars, S.~M. Smith, T.~E. Behrens, S.~Jbabdi,
  \href{https://science.sciencemag.org/content/352/6282/216}{Task-free mri
  predicts individual differences in brain activity during task performance},
  Science 352~(6282) (2016) 216--220.
\newblock \href
  {http://arxiv.org/abs/https://science.sciencemag.org/content/352/6282/216.full.pdf}
  {\path{arXiv:https://science.sciencemag.org/content/352/6282/216.full.pdf}},
  \href {https://doi.org/10.1126/science.aad8127}
  {\path{doi:10.1126/science.aad8127}}.
\newline\urlprefix\url{https://science.sciencemag.org/content/352/6282/216}

\bibitem{Cole16}
M.~S. Cole, T.~Ito, D.~S. Basset, S.~D. H., Activity flow over resting-state
  networks shapes cognitive task activations, Nature Neuroscience 19~(12)
  (2016) 1718--–1726.

\bibitem{Osher19}
D.~E. Osher, J.~A. Brissenden, D.~C. Somers,
  \href{https://doi.org/10.1152/jn.00174.2019}{Predicting an individual’s
  dorsal attention network activity from functional connectivity fingerprints},
  Journal of Neurophysiology 122~(1) (2019) 232--240, pMID: 31066602.
\newblock \href {http://arxiv.org/abs/https://doi.org/10.1152/jn.00174.2019}
  {\path{arXiv:https://doi.org/10.1152/jn.00174.2019}}, \href
  {https://doi.org/10.1152/jn.00174.2019} {\path{doi:10.1152/jn.00174.2019}}.
\newline\urlprefix\url{https://doi.org/10.1152/jn.00174.2019}

\bibitem{Deco12}
G.~Deco, V.~K. Jirsa,
  \href{https://www.jneurosci.org/content/32/10/3366}{Ongoing cortical activity
  at rest: Criticality, multistability, and ghost attractors}, Journal of
  Neuroscience 32~(10) (2012) 3366--3375.
\newblock \href
  {http://arxiv.org/abs/https://www.jneurosci.org/content/32/10/3366.full.pdf}
  {\path{arXiv:https://www.jneurosci.org/content/32/10/3366.full.pdf}}, \href
  {https://doi.org/10.1523/JNEUROSCI.2523-11.2012}
  {\path{doi:10.1523/JNEUROSCI.2523-11.2012}}.
\newline\urlprefix\url{https://www.jneurosci.org/content/32/10/3366}

\bibitem{66}
G.~Deco, A.~Ponce-Alvarez, P.~Hagmann, G.~Romani, D.~Mantini, M.~Corbetta, How
  local excitation-inhibition ratio impacts the whole brain dynamics, Journal
  of Neuroscience 34~(23) (2014) 7886--7898.
\newblock \href {https://doi.org/10.1523/JNEUROSCI.5068-13.2014}
  {\path{doi:10.1523/JNEUROSCI.5068-13.2014}}.

\bibitem{Senden16}
M.~Senden, N.~Reuter, M.~P. van~den Heuvel, R.~Goebel, G.~Deco,
  \href{http://www.sciencedirect.com/science/article/pii/S1053811916306000}{Cortical
  rich club regions can organize state-dependent functional network formation
  by engaging in oscillatory behavior}, NeuroImage 146 (2017) 561--574.
\newblock \href
  {https://doi.org/https://doi.org/10.1016/j.neuroimage.2016.10.044}
  {\path{doi:https://doi.org/10.1016/j.neuroimage.2016.10.044}}.
\newline\urlprefix\url{http://www.sciencedirect.com/science/article/pii/S1053811916306000}

\bibitem{BP03}
J.~Beggs, D.~Plenz, Neuronal avalanches in neocortical circuits, J.
  Neuroscience 23 (2003) 11167.

\bibitem{Fried}
N.~Friedman, S.~Ito, B.~Brinkman, M.~Shimono, R.~Deville, K.~Dahmen, J.~Beggs,
  T.~Butler, Universal critical dynamics in high resolution neuronal avalanche
  data, Physical Review Letters 108~(20) (2012).
\newblock \href {https://doi.org/10.1103/PhysRevLett.108.208102}
  {\path{doi:10.1103/PhysRevLett.108.208102}}.

\bibitem{Shew}
W.~Shew, W.~Clawson, J.~Pobst, Y.~Karimipanah, N.~Wright, R.~Wessel, Adaptation
  to sensory input tunes visual cortex to criticality, Nature Physics 11~(8)
  (2015) 659--663.
\newblock \href {https://doi.org/10.1038/nphys3370}
  {\path{doi:10.1038/nphys3370}}.

\bibitem{Yag}
M.~Yaghoubi, T.~{De Graaf}, J.~Orlandi, F.~Girotto, M.~Colicos, J.~Davidsen,
  Neuronal avalanche dynamics indicates different universality classes in
  neuronal cultures, Scientific Reports 8~(1) (2018).
\newblock \href {https://doi.org/10.1038/s41598-018-21730-1}
  {\path{doi:10.1038/s41598-018-21730-1}}.

\bibitem{brainexp}
J.~Palva, A.~Zhigalov, J.~Hirvonen, O.~Korhonen, K.~Linkenkaer-Hansen,
  S.~Palva, Neuronal long-range temporal correlations and avalanche dynamics
  are correlated with behavioral scaling laws, Proceedings of the National
  Academy of Sciences of the United States of America 110~(9) (2013)
  3585--3590.
\newblock \href {https://doi.org/10.1073/pnas.1216855110}
  {\path{doi:10.1073/pnas.1216855110}}.

\bibitem{Larr}
D.~B. Larremore, W.~L. Shew, J.~G. Restrepo,
  \href{https://link.aps.org/doi/10.1103/PhysRevLett.106.058101}{Predicting
  criticality and dynamic range in complex networks: Effects of topology},
  Phys. Rev. Lett. 106 (2011) 058101.
\newblock \href {https://doi.org/10.1103/PhysRevLett.106.058101}
  {\path{doi:10.1103/PhysRevLett.106.058101}}.
\newline\urlprefix\url{https://link.aps.org/doi/10.1103/PhysRevLett.106.058101}

\bibitem{KC}
O.~Kinouchi, M.~Copelli, Optimal dynamical range of excitable networks at
  criticality, Nature Physics 2~(5) (2006) 348--352.
\newblock \href {https://doi.org/10.1038/nphys289}
  {\path{doi:10.1038/nphys289}}.

\bibitem{stas-bak}
D.~Stassinopoulos, P.~Bak, Democratic reinforcement: A principle for brain
  function, Physical Review E 51~(5) (1995) 5033--5039.
\newblock \href {https://doi.org/10.1103/PhysRevE.51.5033}
  {\path{doi:10.1103/PhysRevE.51.5033}}.

\bibitem{pruessner}
G.~Pruessner, Self-organised criticality: Theory, models and characterisation,
  2012.
\newblock \href {https://doi.org/10.1017/CBO9780511977671}
  {\path{doi:10.1017/CBO9780511977671}}.

\bibitem{MM}
P.~Moretti, M.~A. Mu{\~n}oz,
  \href{http://www.nature.com/doifinder/10.1038/ncomms3521}{{Griffiths phases
  and the stretching of criticality in brain networks}}, Nature Communications
  4 (2013) 2521.
\newblock \href {https://doi.org/10.1038/ncomms3521}
  {\path{doi:10.1038/ncomms3521}}.
\newline\urlprefix\url{http://www.nature.com/doifinder/10.1038/ncomms3521}

\bibitem{HMNcikk}
G.~{\'O}dor, R.~Dickman, G.~{\'O}dor,
  \href{http://www.nature.com/doifinder/10.1038/srep14451}{{Griffiths phases
  and localization in hierarchical modular networks}}, Scientific Reports 5
  (2015) 14451.
\newline\urlprefix\url{http://www.nature.com/doifinder/10.1038/srep14451}

\bibitem{Vojta2006b}
T.~Vojta,
  \href{http://stacks.iop.org/0305-4470/39/i=22/a=R01?key=crossref.efb7d5f24dd1b89df653af8cd0da140b}{{Rare
  region effects at classical, quantum and nonequilibrium phase transitions}},
  Journal of Physics A: Mathematical and General 39 (2006) R143--R205.
\newblock \href {https://doi.org/10.1088/0305-4470/39/22/R01}
  {\path{doi:10.1088/0305-4470/39/22/R01}}.
\newline\urlprefix\url{http://stacks.iop.org/0305-4470/39/i=22/a=R01?key=crossref.efb7d5f24dd1b89df653af8cd0da140b}

\bibitem{Griffiths}
R.~B. Griffiths,
  \href{http://link.aps.org/doi/10.1103/PhysRevLett.23.17}{{Nonanalytic
  Behavior Above the Critical Point in a Random Ising Ferromagnet}}, Phys. Rev.
  Lett. 23~(1) (1969) 17--19.
\newblock \href {https://doi.org/10.1103/PhysRevLett.23.17}
  {\path{doi:10.1103/PhysRevLett.23.17}}.
\newline\urlprefix\url{http://link.aps.org/doi/10.1103/PhysRevLett.23.17}

\bibitem{burstcikk}
G.~{\'O}dor, Slow, bursty dynamics as a consequence of quenched network
  topologies, Phys. Rev. E 89 (2004) 042102.

\bibitem{Johnson}
J.~J. Johnson~S., Torres, J.~Marro, Robust short-term memory without synaptic
  learning, PLoS ONE 8 (2013) e50276.

\bibitem{PSM16}
Y.~Penn, M.~Segal, E.~Moses, Network synchronization in hippocampal neurons,
  Proceedings of the National Academy of Sciences of the United States of
  America 113~(12) (2016) 3341--3346.
\newblock \href {https://doi.org/10.1073/pnas.1515105113}
  {\path{doi:10.1073/pnas.1515105113}}.

\bibitem{MunPNAS}
S.~{Di Santo}, P.~Villegas, R.~Burioni, M.~Mu{\~n}oz, Landau--ginzburg theory
  of cortex dynamics: Scale-free avalanches emerge at the edge of
  synchronization, Proceedings of the National Academy of Sciences of the
  United States of America 115~(7) (2018) E1356--E1365.
\newblock \href {https://doi.org/10.1073/pnas.1712989115}
  {\path{doi:10.1073/pnas.1712989115}}.

\bibitem{Freyer6353}
F.~Freyer, J.~A. Roberts, R.~Becker, P.~A. Robinson, P.~Ritter, M.~Breakspear,
  \href{https://www.jneurosci.org/content/31/17/6353}{Biophysical mechanisms of
  multistability in resting-state cortical rhythms}, Journal of Neuroscience
  31~(17) (2011) 6353--6361.
\newblock \href
  {http://arxiv.org/abs/https://www.jneurosci.org/content/31/17/6353.full.pdf}
  {\path{arXiv:https://www.jneurosci.org/content/31/17/6353.full.pdf}}, \href
  {https://doi.org/10.1523/JNEUROSCI.6693-10.2011}
  {\path{doi:10.1523/JNEUROSCI.6693-10.2011}}.
\newline\urlprefix\url{https://www.jneurosci.org/content/31/17/6353}

\bibitem{DKJR}
G.~Deco, M.~Kringelbach, V.~Jirsa, P.~Ritter, The dynamics of resting
  fluctuations in the brain: Metastability and its dynamical cortical core,
  Scientific Reports 7~(1) (2017).
\newblock \href {https://doi.org/10.1038/s41598-017-03073-5}
  {\path{doi:10.1038/s41598-017-03073-5}}.

\bibitem{DECO2017538}
G.~Deco, J.~Cabral, M.~W. Woolrich, A.~B. Stevner, T.~J. van Hartevelt, M.~L.
  Kringelbach,
  \href{http://www.sciencedirect.com/science/article/pii/S105381191730232X}{Single
  or multiple frequency generators in on-going brain activity: A mechanistic
  whole-brain model of empirical meg data}, NeuroImage 152 (2017) 538 -- 550.
\newblock \href
  {https://doi.org/https://doi.org/10.1016/j.neuroimage.2017.03.023}
  {\path{doi:https://doi.org/10.1016/j.neuroimage.2017.03.023}}.
\newline\urlprefix\url{http://www.sciencedirect.com/science/article/pii/S105381191730232X}

\bibitem{PolRos15}
A.~Politi, M.~Rosenblum, Equivalence of phase-oscillator and integrate-and-fire
  models, Physical Review E - Statistical, Nonlinear, and Soft Matter Physics
  91~(4) (2015).
\newblock \href {https://doi.org/10.1103/PhysRevE.91.042916}
  {\path{doi:10.1103/PhysRevE.91.042916}}.

\bibitem{kura}
Y.~Kuramoto, Chemical Oscillations, Waves, and Turbulence, Springer Series in
  Synergetics, Springer Berlin Heidelberg, 2012.

\bibitem{Acebron}
J.~Acebr{\'o}n, L.~Bonilla, C.~Vicente, F.~Ritort, R.~Spigler, The kuramoto
  model: A simple paradigm for synchronization phenomena, Reviews of Modern
  Physics 77~(1) (2005) 137--185.
\newblock \href {https://doi.org/10.1103/RevModPhys.77.137}
  {\path{doi:10.1103/RevModPhys.77.137}}.

\bibitem{KurCC}
G.~{\'O}dor, J.~Kelling, \href{https://arxiv.org/abs/1903.00385}{Critical
  synchronization dynamics of the kuramoto model on connectome and small world
  graphs}, Scientific Reports 9 (2019) 19621.
\newline\urlprefix\url{https://arxiv.org/abs/1903.00385}

\bibitem{Frus}
P.~Villegas, P.~Moretti, M.~Mu{\~n}oz, Frustrated hierarchical synchronization
  and emergent complexity in the human connectome network, Scientific Reports 4
  (2014).
\newblock \href {https://doi.org/10.1038/srep05990}
  {\path{doi:10.1038/srep05990}}.

\bibitem{chimera}
D.~M. Abrams, S.~H. Strogatz,
  \href{https://link.aps.org/doi/10.1103/PhysRevLett.93.174102}{Chimera states
  for coupled oscillators}, Phys. Rev. Lett. 93 (2004) 174102.
\newblock \href {https://doi.org/10.1103/PhysRevLett.93.174102}
  {\path{doi:10.1103/PhysRevLett.93.174102}}.
\newline\urlprefix\url{https://link.aps.org/doi/10.1103/PhysRevLett.93.174102}

\bibitem{Frus-noise}
P.~Villegas, J.~Hidalgo, P.~Moretti, M.~Mu{\~n}oz, Complex synchronization
  patterns in the human connectome network, 2016, pp. 69--80.
\newblock \href {https://doi.org/10.1007/978-3-319-29228-1_7}
  {\path{doi:10.1007/978-3-319-29228-1_7}}.

\bibitem{FrusB}
A.~Mill{\'a}n, J.~Torres, G.~Bianconi, Complex network geometry and frustrated
  synchronization, Scientific Reports 8~(1) (2018).
\newblock \href {https://doi.org/10.1038/s41598-018-28236-w}
  {\path{doi:10.1038/s41598-018-28236-w}}.

\bibitem{Homeo-inh}
M.~Remme, W.~Wadman, Homeostatic scaling of excitability in recurrent neural
  networks, PLoS Computational Biology 8~(5) (2012).
\newblock \href {https://doi.org/10.1371/journal.pcbi.1002494}
  {\path{doi:10.1371/journal.pcbi.1002494}}.

\bibitem{65}
F.~Droste, A.-L. Do, T.~Gross, Analytical investigation of self-organized
  criticality in neural networks, Journal of the Royal Society Interface
  10~(78) (2013).
\newblock \href {https://doi.org/10.1098/rsif.2012.0558}
  {\path{doi:10.1098/rsif.2012.0558}}.

\bibitem{67}
P.~Hellyer, B.~Jachs, C.~Clopath, R.~Leech, Local inhibitory plasticity tunes
  macroscopic brain dynamics and allows the emergence of functional brain
  networks, NeuroImage 124 (2016) 85--95.
\newblock \href {https://doi.org/10.1016/j.neuroimage.2015.08.069}
  {\path{doi:10.1016/j.neuroimage.2015.08.069}}.

\bibitem{68}
P.~Hellyer, C.~Clopath, A.~Kehagia, F.~Turkheimer, R.~Leech, From homeostasis
  to behavior: Balanced activity in an exploration of embodied dynamic
  environmental-neural interaction, PLoS computational biology 13~(8) (2017)
  e1005721.
\newblock \href {https://doi.org/10.1371/journal.pcbi.1005721}
  {\path{doi:10.1371/journal.pcbi.1005721}}.

\bibitem{CCdyncikk}
G.~{\'O}dor,
  \href{https://link.aps.org/doi/10.1103/PhysRevE.94.062411}{{Critical dynamics
  on a large human Open Connectome network}}, Phys. Rev. E 94~(6) (2016)
  062411.
\newline\urlprefix\url{https://link.aps.org/doi/10.1103/PhysRevE.94.062411}

\bibitem{Rocha2008}
R.~Rocha, L.~Ko{\c c}illari, S.~Suweis, M.~Corbetta, A.~Maritan, Homeostatic
  plasticity and emergence of functional networks in a whole-brain model at
  criticality, Scientific Reports 8~(1) (2018).
\newblock \href {https://doi.org/10.1038/s41598-018-33923-9}
  {\path{doi:10.1038/s41598-018-33923-9}}.

\bibitem{CCcikk}
M.~Gastner, G.~{\'O}dor, The topology of large open connectome networks for the
  human brain, Scientific Reports 6 (2016) 27249.
\newblock \href {https://doi.org/10.1038/srep27249}
  {\path{doi:10.1038/srep27249}}.

\bibitem{10.1143/PTP.79.39}
H.~Sakaguchi, \href{https://doi.org/10.1143/PTP.79.39}{{Cooperative Phenomena
  in Coupled Oscillator Systems under External Fields}}, Progress of
  Theoretical Physics 79~(1) (1988) 39--46.
\newblock \href
  {http://arxiv.org/abs/http://oup.prod.sis.lan/ptp/article-pdf/79/1/39/6867039/79-1-39.pdf}
  {\path{arXiv:http://oup.prod.sis.lan/ptp/article-pdf/79/1/39/6867039/79-1-39.pdf}},
  \href {https://doi.org/10.1143/PTP.79.39} {\path{doi:10.1143/PTP.79.39}}.
\newline\urlprefix\url{https://doi.org/10.1143/PTP.79.39}

\bibitem{OCP}
\href{https://neurodata.io}{Neurodata} (2015).
\newline\urlprefix\url{https://neurodata.io}

\bibitem{NumR}
W.~Press, S.~Teukolsky, W.~Vetterling, B.~Flannery,
  \href{http://numerical.recipes}{Numerical Recipes 3rd Edition: The Art of
  Scientific Computing}, Cambridge University Press, 2007.
\newline\urlprefix\url{http://numerical.recipes}

\bibitem{boostOdeInt}
K.~Ahnert, M.~Mulansky, \href{https://odeint.com}{Boost::odeint}.
\newline\urlprefix\url{https://odeint.com}

\bibitem{GPU-kur}
J.~Kelling, G.~{\'O}dor, S.~Gemming, to be published (2020).

\bibitem{DTI}
B.~Landman, A.~Huang, A.~Gifford, D.~Vikram, I.~Lim, J.~Farrell, J.~Bogovic,
  J.~Hua, M.~Chen, S.~Jarso, S.~Smith, S.~Joel, S.~Mori, J.~Pekar, P.~Barker,
  J.~Prince, P.~van Zijl, Multi-parametric neuroimaging reproducibility: A 3-t
  resource study, NeuroImage 54~(4) (2011) 2854--2866.
\newblock \href {https://doi.org/10.1016/j.neuroimage.2010.11.047}
  {\path{doi:10.1016/j.neuroimage.2010.11.047}}.

\bibitem{MIG}
W.~{Gray Roncal}, Z.~H. {Koterba}, D.~{Mhembere}, D.~M. {Kleissas}, J.~T.
  {Vogelstein}, R.~{Burns}, A.~R. {Bowles}, D.~K. {Donavos}, S.~{Ryman}, R.~E.
  {Jung}, L.~{Wu}, V.~{Calhoun}, R.~J. {Vogelstein}, Migraine: Mri graph
  reliability analysis and inference for connectomics, in: 2013 IEEE Global
  Conference on Signal and Information Processing, 2013, pp. 313--316.
\newblock \href {https://doi.org/10.1109/GlobalSIP.2013.6736878}
  {\path{doi:10.1109/GlobalSIP.2013.6736878}}.

\bibitem{TraagWaltmaVanEck2018_LouvainLeiden}
V.~A. Traag, L.~Waltman, N.~J. van Eck,
  \href{https://doi.org/10.1038/s41598-019-41695-z}{From louvain to leiden:
  guaranteeing well-connected communities}, Scientific Reports 9~(1) (2019)
  5233.
\newblock \href {https://doi.org/10.1038/s41598-019-41695-z}
  {\path{doi:10.1038/s41598-019-41695-z}}.
\newline\urlprefix\url{https://doi.org/10.1038/s41598-019-41695-z}

\bibitem{igraph}
\href{https://igraph.org/}{igraph}.
\newline\urlprefix\url{https://igraph.org/}

\bibitem{newcomp}
C.~Delettre, A.~Messé, L.-A. Dell, O.~Foubet, K.~Heuer, B.~Larrat, S.~Meriaux,
  J.-F. Mangin, I.~Reillo, C.~de~Juan~Romero, V.~Borrell, R.~Toro, C.~C.
  Hilgetag, \href{https://doi.org/10.1162/netn_a_00098}{Comparison between
  diffusion mri tractography and histological tract-tracing of cortico-cortical
  structural connectivity in the ferret brain}, Network Neuroscience 3~(4)
  (2019) 1038--1050.
\newblock \href {http://arxiv.org/abs/https://doi.org/10.1162/netn_a_00098}
  {\path{arXiv:https://doi.org/10.1162/netn_a_00098}}, \href
  {https://doi.org/10.1162/netn\_a\_00098} {\path{doi:10.1162/netn\_a\_00098}}.
\newline\urlprefix\url{https://doi.org/10.1162/netn_a_00098}

\bibitem{KH}
M.~Kaiser, C.~Hilgetag, Optimal hierarchical modular topologies for producing
  limited sustained activation of neural networks, Frontiers in
  Neuroinformatics 4 (2010).
\newblock \href {https://doi.org/10.3389/fninf.2010.00008}
  {\path{doi:10.3389/fninf.2010.00008}}.

\bibitem{Ponce-Deco-Plos15}
A.~Ponce-Alvarez, Deco, P.~Hagmann, G.~Romani, Mantini, D.~Corbetta,
  Resting-state temporal synchronization networks emerge from connectivity
  topology and heterogeneity, PLoSComput Biol 11~(2) (2015).
\newblock \href {https://doi.org/10.1371/journal.pcbi.1004100}
  {\path{doi:10.1371/journal.pcbi.1004100}}.

\bibitem{MS90}
M.~P. C., S.~S. H., Phase diagram for the collective bahavior of limit-cycle
  oscillators, Phys. Rev. Lett. 65~(14) (1990) 1701--1704.

\bibitem{PhysRevE.72.046211}
D.~Paz\'o,
  \href{https://link.aps.org/doi/10.1103/PhysRevE.72.046211}{Thermodynamic
  limit of the first-order phase transition in the kuramoto model}, Phys. Rev.
  E 72 (2005) 046211.
\newblock \href {https://doi.org/10.1103/PhysRevE.72.046211}
  {\path{doi:10.1103/PhysRevE.72.046211}}.
\newline\urlprefix\url{https://link.aps.org/doi/10.1103/PhysRevE.72.046211}

\bibitem{Feng_2015}
Y.-E. Feng, H.-H. Li,
  \href{https://doi.org/10.1088%2F0256-307x%2F32%2F6%2F060502}{The dependence
  of chimera states on initial conditions}, Chinese Physics Letters 32~(6)
  (2015) 060502.
\newblock \href {https://doi.org/10.1088/0256-307x/32/6/060502}
  {\path{doi:10.1088/0256-307x/32/6/060502}}.
\newline\urlprefix\url{https://doi.org/10.1088%2F0256-307x%2F32%2F6%2F060502}

\bibitem{PhysRevLett.101.264103}
A.~Pikovsky, M.~Rosenblum,
  \href{https://link.aps.org/doi/10.1103/PhysRevLett.101.264103}{Partially
  integrable dynamics of hierarchical populations of coupled oscillators},
  Phys. Rev. Lett. 101 (2008) 264103.
\newblock \href {https://doi.org/10.1103/PhysRevLett.101.264103}
  {\path{doi:10.1103/PhysRevLett.101.264103}}.
\newline\urlprefix\url{https://link.aps.org/doi/10.1103/PhysRevLett.101.264103}

\bibitem{LAING20091569}
C.~R. Laing,
  \href{http://www.sciencedirect.com/science/article/pii/S0167278909001377}{The
  dynamics of chimera states in heterogeneous kuramoto networks}, Physica D:
  Nonlinear Phenomena 238~(16) (2009) 1569 -- 1588.
\newblock \href {https://doi.org/https://doi.org/10.1016/j.physd.2009.04.012}
  {\path{doi:https://doi.org/10.1016/j.physd.2009.04.012}}.
\newline\urlprefix\url{http://www.sciencedirect.com/science/article/pii/S0167278909001377}

\bibitem{PhysRevE.89.022914}
Y.~Zhu, Z.~Zheng, J.~Yang,
  \href{https://link.aps.org/doi/10.1103/PhysRevE.89.022914}{Chimera states on
  complex networks}, Phys. Rev. E 89 (2014) 022914.
\newblock \href {https://doi.org/10.1103/PhysRevE.89.022914}
  {\path{doi:10.1103/PhysRevE.89.022914}}.
\newline\urlprefix\url{https://link.aps.org/doi/10.1103/PhysRevE.89.022914}

\bibitem{CABRAL2011130}
J.~Cabral, E.~Hugues, O.~Sporns, G.~Deco,
  \href{http://www.sciencedirect.com/science/article/pii/S1053811911003880}{Role
  of local network oscillations in resting-state functional connectivity},
  NeuroImage 57~(1) (2011) 130 -- 139.
\newblock \href
  {https://doi.org/https://doi.org/10.1016/j.neuroimage.2011.04.010}
  {\path{doi:https://doi.org/10.1016/j.neuroimage.2011.04.010}}.
\newline\urlprefix\url{http://www.sciencedirect.com/science/article/pii/S1053811911003880}

\bibitem{POWcikk}
G.~{\'O}dor, B.~Hartmann, Heterogeneity effects in power grid network models,
  Physical Review E 98~(2) (2018).
\newblock \href {https://doi.org/10.1103/PhysRevE.98.022305}
  {\path{doi:10.1103/PhysRevE.98.022305}}.

\bibitem{BU08}
L.~Basnarkov, V.~Urumov,
  \href{https://link.aps.org/doi/10.1103/PhysRevE.78.011113}{Kuramoto model
  with asymmetric distribution of natural frequencies}, Phys. Rev. E 78 (2008)
  011113.
\newblock \href {https://doi.org/10.1103/PhysRevE.78.011113}
  {\path{doi:10.1103/PhysRevE.78.011113}}.
\newline\urlprefix\url{https://link.aps.org/doi/10.1103/PhysRevE.78.011113}

\bibitem{Tau-pet-10}
G.~Tau, B.~Peterson, Normal development of brain circuits, Neuropsychopharmacol
  35 (2010) 147–168.

\bibitem{odorbook}
G.~{\'O}dor, Universality in nonequilibrium lattice systems: Theoretical
  foundations, 2008.
\newblock \href {https://doi.org/10.1142/6813} {\path{doi:10.1142/6813}}.

\bibitem{PELISSETTO2002549}
A.~Pelissetto, E.~Vicari,
  \href{http://www.sciencedirect.com/science/article/pii/S0370157302002193}{Critical
  phenomena and renormalization-group theory}, Physics Reports 368~(6) (2002)
  549 -- 727.
\newblock \href {https://doi.org/https://doi.org/10.1016/S0370-1573(02)00219-3}
  {\path{doi:https://doi.org/10.1016/S0370-1573(02)00219-3}}.
\newline\urlprefix\url{http://www.sciencedirect.com/science/article/pii/S0370157302002193}

\bibitem{3DXYZ}
L.~M. Jensen, B.~J. Kim, P.~Minnhagen, Dynamic critical behaviors of
  three-dimensional xy models related to superconductors/superfluids, Europhys.
  Lett. 49 (2000) 644 -- 650.

\end{thebibliography}
%%%%%%%%%%%%%%%%%%%%%%%%%%%%%%%%%%%%%%%%%%%%%%%%%%%%%%%%%%%%%%%%%%%%%%%%

\end{document}